# Single photon production in $p\bar{p}$ collision in the collinear and $k_t$-factorisation framework and investigation of their differences

M. Mousavi

Department of Physics, University of Tehran, 14395547, Tehran, Iran

**Abstract**

The cross section of the single photon in $p\bar{p}$ collision is calculated. There are many subprocesses, for example the LO and NLO collisions, leading to the single photon. The LO collisions, the one of which (Compton scattering) is strongly dependent on the gluon distribution function, are investigated. Both the off-shell ($k_t$ dependent) and on-shell partons are considered in the diffusions, and their differences are discussed. The unintegrated parton distribution functions (UPDFs), used for the off-shell partons, are calculated by the Martin-Ryskin-Watt (MRW) method. In the case of the on-shell partons, the integrated parton distribution functions (PDFs), are obtained by MMHT 2014 LO PDFs which are determined experimentally. To investigate the accuracy of the theoretical results, they are compared with the experimental data that are taken by the DØ and CDF collaborations at Fermilab Tevatron.

## 1- Introduction

The study of the single photon production in the collision of hadrons has been an interesting experimental [1-17] and theoretical [18-49] subject since the past. It provides knowledge of the structure of hadrons, moreover, is a proof for the hard subprocesses occurring behind the collision of two hadrons. Conventionally, in hard processes, the incoming partons are considered the on-shell, which only carry longitudinal momentum fraction $x$ of their parent hadrons and consequently are parallel to them. Their parton distribution function is the function of $x$ and $\mu^2$, $f(x, \mu^2)$. It corresponds to the density of partons in a hadron with longitudinal momentum fraction $x$, integrated over the parton transverse momentum up to $k_t = \mu$. It satisfies DGLAP evolution in the factorization scale $\mu$, moreover, it is determined from the global analyses of DIS and high energy collisions. Today, the incoming partons are assumed to be off-shell, which depends on the transverse momentum ($k_t$ dependent). The dependence on transverse momentum is a very important difference between off-shell and on-shell partons. It leads the unintegrated parton distribution function to be calculated in a more complicated way then PDFs and a function of $x$, $k_t^2$, and $\mu^2$, $f(x, k_t^2, \mu^2)$. The unintegrated parton distribution function (UPDF) corresponds to the number of the partons with longitudinal momentum fraction $x$ and transverse momentum $k_t^2$ in a proton. The purpose of the study of the cross section of the single photon in two off-shell and on-shell states is to understand the differences between off-shell and on-shell partons which will appear at small $x$ (or high energies $x \sim 1/\sqrt{s}$) where the consideration of $k_t$ dependent becomes crucial. The unintegrated parton distribution functions are beneficial because they can accurately explain the hard subprocesses, their Feynman diagrams, and true kinematics at small $x$, even at the LO level.

---

Corresponding author: mohammadmousavi1@ut.ac.ir

Many subprocesses, which lead to the single photon, occur in $p\bar{p}$ collision. They are arranged to leading-order (LO), next to leading order (NLO), and other higher levels according to the order of their total amplitude. The on-shell matrix elements satisfy the gauge invariance and in the case of off-shell matrix elements, because of the small $x$ region, they also satisfy the gauge invariance (for more details please see [50]). In addition, the observed single photon may be result of the fragmentation process, where a final state quark or gluon fragments into a photon [22, 51]. The goal is to study the parton distribution functions in $p\bar{p}$ collisions, and the fragmentation process disrupts this aim. Therefore, the constraints are imposed on the experiments and theoretical calculations of the cross sections to reduce the impact of the fragmentation processes. The fragmentation processes in theoretical calculations of the single photon cross section are not considered and two LO direct subprocesses are only calculated. The LO direct collisions, Compton ( q + g → γ + q ) and annihilation scattering ( q + q̄ → γ + g ), are strongly sensitive to the quarks and gluons in hadrons. For $p\bar{p}$ collision, the distribution function of an antiquark in antiproton is equivalent to the distribution function of the same flavor quark in proton, it means $u_{\bar{P}}^{\bar{q}}(x) = u_{P}^{q}(x)$, where $q$ and $\bar{q}$ are the flavours of the quark and antiquark, respectively, $u_{P}^{q}(x)$ is the distribution function of the quark $q$ in the proton, and $u_{\bar{P}}^{\bar{q}}(x)$ is the distribution function of the antiquark $\bar{q}$ in the antiproton. This equality is one of the reasons which leads the cross section of the annihilation processes to become more dominant than the Compton processes at large $x$.

In this paper, it is used the integrated parton distribution functions (PDFs) which are experimentally calculated by Martin-Motylinski-Harland-Thorne (MMHT) 2014 [52]. There are some methods to calculate the unintegrated parton distribution functions [53-72] and attempts [73-79] to investigate their validity. The Balitsky-Fadin-Kuraev-Lipatov (BFKL) prescription [59-61] gives a better answer to UPDFs at small $x$ where the large terms of the BFKL equation are proportional to $\ln(1/x)$. The Dokshitzer–Gribov–Lipatov–Altarelli–Parisi (DGLAP) equation [53-56], which is proportional to $\ln(\mu^2)$, obtains the better answer at large $x$. At [63], the authors give a method that unified the BFKL and DGLAP evelutions, consequently it is valid for all ranges of $x$. Moreover, they give another method, Kimber-Martin-Ryskin (KMR), which is started by pure DGLAP equation and bring the dependence on the scale $\mu$ at the last evolution step. The KMR method has very similar results with the unified one and is simpler than it. Therefore, it is valid for all ranges of $x$. The strong $k_t$ ordering ($\ldots \ll k_{n-1,t} \ll k_{n,t} \sim \mu$) in DGLAP evolution ensures the angular ordering that determines the maximum z in the KMR method. To calculate the unintegrated parton distribution functions, it is utilised the newer method, Martin-Ryskin-Watt (MRW), which is similar to KMR but with some differences. First, in the KMR, angular ordering is imposed on both emitted quarks and gluons, while in the MRW, it is only imposed on the gluons because of color coherence of gluons [67, 68]. Second, the KMR approach is based on the $k_t$-factorisation while in the MRW, UPDFs depend on both z and $k_t$, $f(x, z, k_t^2, \mu^2)$.

Another role of scale $\mu$ is acting as the factorization scale which is used to relate $f(x, \mu^2)$ to $f(x, k_t^2, \mu^2)$ as the following relation[1]:

$$a(x, \mu^2) = \int_0^{\mu^2} \frac{dk_t^2}{k_t^2} f_a(x, k_t^2, \mu^2), \qquad (1)$$

Where $a(x,\mu^2) = xq(x,\mu^2)$ or $xg(x,\mu^2)$. At [32], the authors to normalise $f(x,k_t^2,\mu^2)$ used the following different relation:

$$a(x,\mu^2) = \int_0^{\mu^2} dk_t^2\, f_a(x,k_t^2,\mu^2), \tag{2}$$

According to their assertion, since they included $k_t^2$ into transverse momentum dependent parton density functions (TMDs), their normalization relation is different. This difference causes the equation of their cross section to be different from the used equation in this paper. Although they used different normalization, their results are similar to the results of this paper.

In this paper, first, the parton distribution functions and the total amplitude of two LO collisions for both off-shell and on-shell incoming particles are calculated. In section 4, the kinematics and relations to calculate the cross section of the single direct photon will be presented. Finally, in section 5, theoretical results are compared with the data of three experiments that are taken by DØ and CDF Fermilab Tevatron.

## 2- Parton distribution functions

Parton distribution functions correspond to the density (the number of the specific parton in all partons constructing a hadron) of partons in a hadron. At the parton model, for measuring the cross section of the $p\bar{p}$ collision, one must calculate the number of partons contributing in subprocesses then multiply them at the related subprocesses cross section. For example, in Compton scattering, one must find the number of gluons and quarks which come from their parent hadron then multiply them at the cross section of Compton scattering to find the cross section of the $p\bar{p}$ collision. Since the aim is to study both the on-shell and off-shell incoming partons, two states for the density partons are presented

### 2-1 Integrated parton distribution functions

The integrated parton distribution functions (PDFs) satisfies the DGLAP equation at the scale $\mu$:

$$\frac{\partial}{\partial \log(\mu^2)} \begin{pmatrix} q_i(x,\mu^2) \\ g(x,\mu^2) \end{pmatrix} = \frac{\alpha_s(\mu^2)}{2\pi} \sum_{q_j,\bar{q}_j} \int_x^1 \frac{d\zeta}{\zeta} \times \begin{pmatrix} P_{q_i q_j}\left(\frac{x}{\zeta},\alpha_s(t)\right) & P_{q_i g}\left(\frac{x}{\zeta},\alpha_s(t)\right) \\ P_{g q_j}\left(\frac{x}{\zeta},\alpha_s(t)\right) & P_{gg}\left(\frac{x}{\zeta},\alpha_s(t)\right) \end{pmatrix} \begin{pmatrix} q_j(x,t) \\ g(x,t) \end{pmatrix} \tag{3}$$

The above equation is a $(2n_f + 1)$ dimensional matrix, $n_f$ is the number of active flavours.

---

1- The equality of relation (1) is not very rigorous, at [63] the authors impose $c^2\mu^2$ ($c = 0.6 - 0.8$) to the up limit of integral to make it more precise

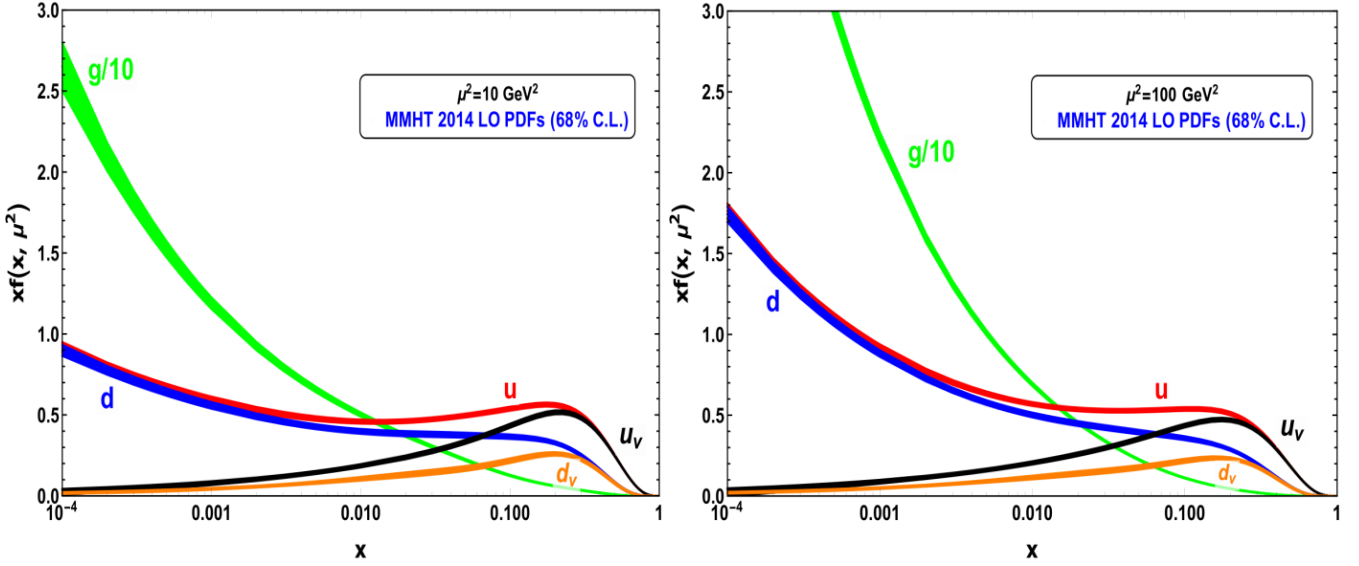

Figure 1: The LO PDFs according to MMHT 2014 at $\mu^2 = 10$ GeV$^2$ (left plot) and $\mu^2 = 100$ GeV$^2$ (right plot), with associated 68% confidence-level uncertainty bands. The error bands are calculated according to Equ. (51) and (52) in [82]

In Equ. (3), $q_i$ is the quark with flavor $i$, $\bar{q}_i$ is the antiquark with flavor $i$, $q_i(x, \mu^2)$ is the distribution function of the quark with flavor $i$, $g(x, \mu^2)$ is the gluon distribution function. The function $P_{ab}(z, \alpha_s(t))$ ($a$ & $b$ are the quarks and gluon) is the splitting function which has a perturbative expansion in running coupling constant:

$$P_{q_i q_j}(z, \alpha_s) = \delta_{ij} P^0_{q_i q_j}(z) + \frac{\alpha_s}{2\pi} P^1_{q_i q_j}(z, \alpha_s) + \cdots \quad (4)$$

$$P_{qg}(z, \alpha_s) = P^0_{qg}(z) + \frac{\alpha_s}{2\pi} P^1_{qg}(z, \alpha_s) + \cdots \quad (5)$$

$$P_{gq}(z, \alpha_s) = P^0_{gq}(z) + \frac{\alpha_s}{2\pi} P^1_{gq}(z, \alpha_s) + \cdots \quad (6)$$

$$P_{gg}(z, \alpha_s) = P^0_{gg}(z) + \frac{\alpha_s}{2\pi} P^1_{gg}(z, \alpha_s) + \cdots \quad (7)$$

The functions $P^0_{ab}(z)$ and $P^1_{ab}(z, \alpha_s)$ are the LO and NLO splitting functions [55, 80, 81], respectively. Both the splitting functions and running coupling constant are perturbative. Consequently, if one takes the first items of equation (4-7) and LO coupling constant, then put them in equation (3), the calculated PDFs will be LO. The integrated parton distribution functions (PDFs) are determined experimentally [52, 82]. In all calculations of this paper, the PDFs are obtained by MMHT 2014 LO PDFs [49]. The functions $a(x, \mu^2)$ are plotted in figure 1 for some quarks and gluon with respect to $x$. They are determined by the LO MMHT 2014 with 68% confidence-level uncertainty bands for $\mu^2 = 10$ GeV$^2$ and $\mu^2 = 100$ GeV$^2$.

## 2-2 Unintegrated parton distribution functions

The number of partons "a" in the proton with the momentum fraction between $x$ and $x + dx$ and transverse momentum $k_t$ between zero and the factorisation $\mu$ is:

$$a(x, \mu^2) \frac{dx}{x}, \tag{8}$$

While, for off-shell particles, the number of partons "a" with the momentum fraction between $x$ and $x + dx$ and transverse momentum $k_t^2$ between $k_t^2 + dk_t^2$ is:

$$f_a(x, k_t^2, \mu^2) \frac{dx}{x} \frac{dk_t^2}{k_t^2}, \tag{9}$$

By equality of (8) and (9), we find the following normalization relation:

$$a(x, \mu^2) = \int_0^{\mu^2} \frac{dk_t^2}{k_t^2} f_a(x, k_t^2, \mu^2). \tag{10}$$

To calculate the function $f_a(x, k_t^2, \mu^2)$, we use the MRW method [67, 68]. This method is similar to the KMR [63] but with two differences that refine and extend the KMR last step procedure for determining the UPDFs. The first one is; in the KMR, angular ordering was imposed on both emitted quarks and gluons while in the MRW, it is only used for gluons. The second is; the KMR is based on $k_t$- factorisation and partons have virtuality $-k_t^2$, but in the MRW, the authors extend those assumptions and define virtuality $-k_t^2/(1-z)$ which leads the UPDFs also become a function of z, $f_a(x, z, k_t^2, \mu^2)$.

In the MRW method, the procedure of calculating UPDFs is started by the LO DGLAP equation evaluated at a scale $k_t$:

$$\frac{\partial a(x, k_t^2)}{\partial \log(k_t^2)} = \frac{\alpha_s(k_t^2)}{2\pi} \sum_{b=q,g} \left[ \int_x^1 dz\, P_{ab}(z) b\left(\frac{x}{z}, k_t^2\right) - a(x, k_t^2) \int_0^1 d\zeta\, \zeta\, P_{ba}(\zeta) \right], \tag{11}$$

Where the used items are introduced at (3). At Equ. (11), the first term, which is corresponded to the real parton emission, changes $k_t$ of the partons in the evolution. The second term, which is corresponded to the virtual part of emission, includes all loops that exist in the real parton emission before it goes to the hard subprocess. The virtual parts do not affect the transverse momentum of the emitted partons. The extra factor $\zeta$ in te virtual part avoids double-counting the s-channel and t-channel partons and is equivalent to ½ when integrating over it and summing over $b$.

All loops are resummed to the Sudakov form factor, $T_a(k_t^2, \mu^2)$, which gives the probability of evolving from the value $k_t$ to $\mu$ without the parton emission.

$$T_a(k_t^2, \mu^2) \equiv \exp\left(-\int_{k_t^2}^{\mu^2} \frac{d\kappa_t^2}{\kappa_t^2} \frac{\alpha_s(\kappa_t^2)}{2\pi} \sum_b \int_0^1 d\zeta\, \zeta\, P_{ba}(\zeta)\right), \tag{12}$$

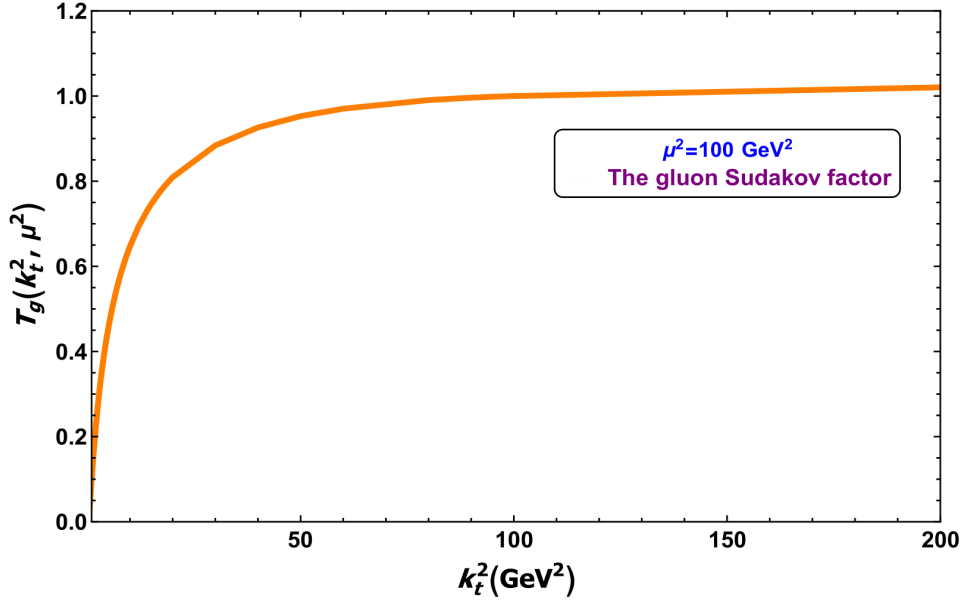

Figure 2: The Sudakov factor for a gluon with $n_f = 5$ and $\Lambda = 0.226$ GeV at $\mu^2 = 100$ GeV².

This factor is defined for $k_t > \mu_0 = 1$ GeV where $\mu_0$ is the minimum scale for which the DGLAP equation is valid. Moreover, when $k_t$ is equal $\mu$, the Sudakov factor becomes one, and for $k_t > \mu$ it will be also constant. In figure 2, the factor is plotted with respect to $k_t^2$ for $\mu^2 = 100$ GeV². The derivative of the Sudakov factor is:

$$\frac{\partial T_a(k_t^2, \mu^2)}{\partial \log k_t^2} = T_a(k_t^2, \mu^2) \frac{\alpha_s(k_t^2)}{2\pi} \sum_b \int_0^1 d\zeta \, \zeta \, P_{ab}(\zeta), \qquad (13)$$

In the MRW method, the following relation is used to find the unintegrated parton distribution functions:

$$f_a(x, k_t^2, \mu^2) = \frac{\partial}{\partial \log k_t^2}[a(x, k_t^2) T_a(k_t^2, \mu^2)] = T_a(k_t^2, \mu^2) \frac{\partial a(x, k_t^2)}{\partial \log k_t^2} + a(x, k_t^2) \frac{\partial T_a(k_t^2, \mu^2)}{\partial \log k_t^2}, (14)$$

Putting (11) and (13) in (14) gives:

$$f_a(x, k_t^2, \mu^2) = \frac{\alpha_s(k_t^2)}{2\pi} T_a(k_t^2, \mu^2) \sum_{b=q,g} \int_x^1 dz \, P_{ab}(z) b\left(\frac{x}{z}, k_t^2\right). \qquad (15)$$

Equ. (15) is also valid for $k_t > \mu_0 = 1$ GeV. By Equ. (14) and (10), UPDFs for $k_t < \mu_0 = 1$ GeV will be determined:

$$\int_{\mu_0^2}^{\mu^2} f_a(x, k_t^2, \mu^2) \frac{dk_t^2}{k_t^2} = [a(x, k_t^2) T_a(k_t^2, \mu^2)]_{k_t=\mu_0}^{k_t=\mu} = a(x, k_t^2) - a(x, \mu_0^2) T_a(\mu_0^2, \mu^2), \quad (16)$$

Then UPDFs for $k_t < 1$ GeV are:

$$\left.\frac{f_a(x, k_t^2, \mu^2)}{k_t^2}\right|_{k_t<\mu_0} = \frac{1}{\mu_0^2} a(x, \mu_0^2) T_a(\mu_0^2, \mu^2), \qquad (17)$$

In Equ. (12) and (15), the splitting functions $P_{qq}(z)$ and $P_{gg}(z)$ have singularities, originated from soft gluon emission, at $z = 1$. These singularities cancel between the real and virtual parts of the DGLAP equation but the Sudakov factor must be regulated. Angular ordering that originated from colour coherence imposes the cut off on splitting fraction $z$ for the splitting functions where a real gluon is emitted in the $s$-channel. In the MRW method, by applying the angular ordering, especially in the last step of evolution, the cut off on $z$ is:

$$z < \frac{\mu}{\mu + k_t}. \tag{19}$$

For other evolutions steps, the strong ordering in the transverse momentum automatically guaranties angular ordering. Moreover, the quarks (fermion) have no coherence effect, consequently, their emission is not limited by angular ordering. (this is one of the differences between the KMR and MRW method). After imposing the restriction on $z$ and using the symmetries $P_{gq}(1 - \zeta) = P_{qq}(\zeta)$ and $P_{qg}(1 - \zeta) = P_{qg}(\zeta)$, Equ. (12) becomes :

$$T_q(k_t^2, \mu^2) \equiv \exp\left(-\int_{k_t^2}^{\mu^2} \frac{d\kappa_t'^2}{\kappa_t'^2} \frac{\alpha_s(\kappa_t'^2)}{2\pi} \int_0^1 d\zeta \, P_{qq}(\zeta) \, \Theta(1 - \Delta - \zeta)\right). \tag{20}$$

And

$$T_g(k_t^2, \mu^2) \equiv \exp\left(-\int_{k_t^2}^{\mu^2} \frac{d\kappa_t'^2}{\kappa_t'^2} \frac{\alpha_s(\kappa_t'^2)}{2\pi} \int_0^1 d\zeta \, (\zeta \, P_{gg}(\zeta) \Theta(1 - \Delta - \zeta) \, \Theta(\zeta - \Delta) + n_f P_{qg}(\zeta))\right). \tag{21}$$

The same acting is done on Equ. (15), therefore the final unintegrated parton distribution functions are :

$$f_q(x, k_t^2, \mu^2) = T_q(k_t^2, \mu^2) \frac{\alpha_s(k_t^2)}{2\pi} \int_x^1 dz \left[P_{qq}(z) \frac{x}{z} q\left(\frac{x}{z}, k_t^2\right) \Theta(1 - \Delta - z) + P_{qg}(z) \frac{x}{z} g\left(\frac{x}{z}, k_t^2\right)\right]. \tag{22}$$

And for gluon

$$f_g(x, k_t^2, \mu^2) = T_g(k_t^2, \mu^2) \frac{\alpha_s(k_t^2)}{2\pi} \int_x^1 dz \left[\sum_q P_{gq}(z) \frac{x}{z} q\left(\frac{x}{z}, k_t^2\right) + P_{gg}(z) \frac{x}{z} g\left(\frac{x}{z}, k_t^2\right) \Theta(1 - \Delta - z)\right]. \tag{23}$$

Where

$$\Delta = \frac{k_t}{k_t + \mu}. \tag{24}$$

In figure 3 and 4, unintegrated parton distribution functions are plotted with respect to $k_t^2$ and $x$, respectively. The plotted functions are calculated by Equ. (20-23) in which MMHT 2014 LO integrated PDFs, LO splitting functions, and LO running coupling constant with $n_f = 5$ and $\Lambda = 0.226$ GeV are used.

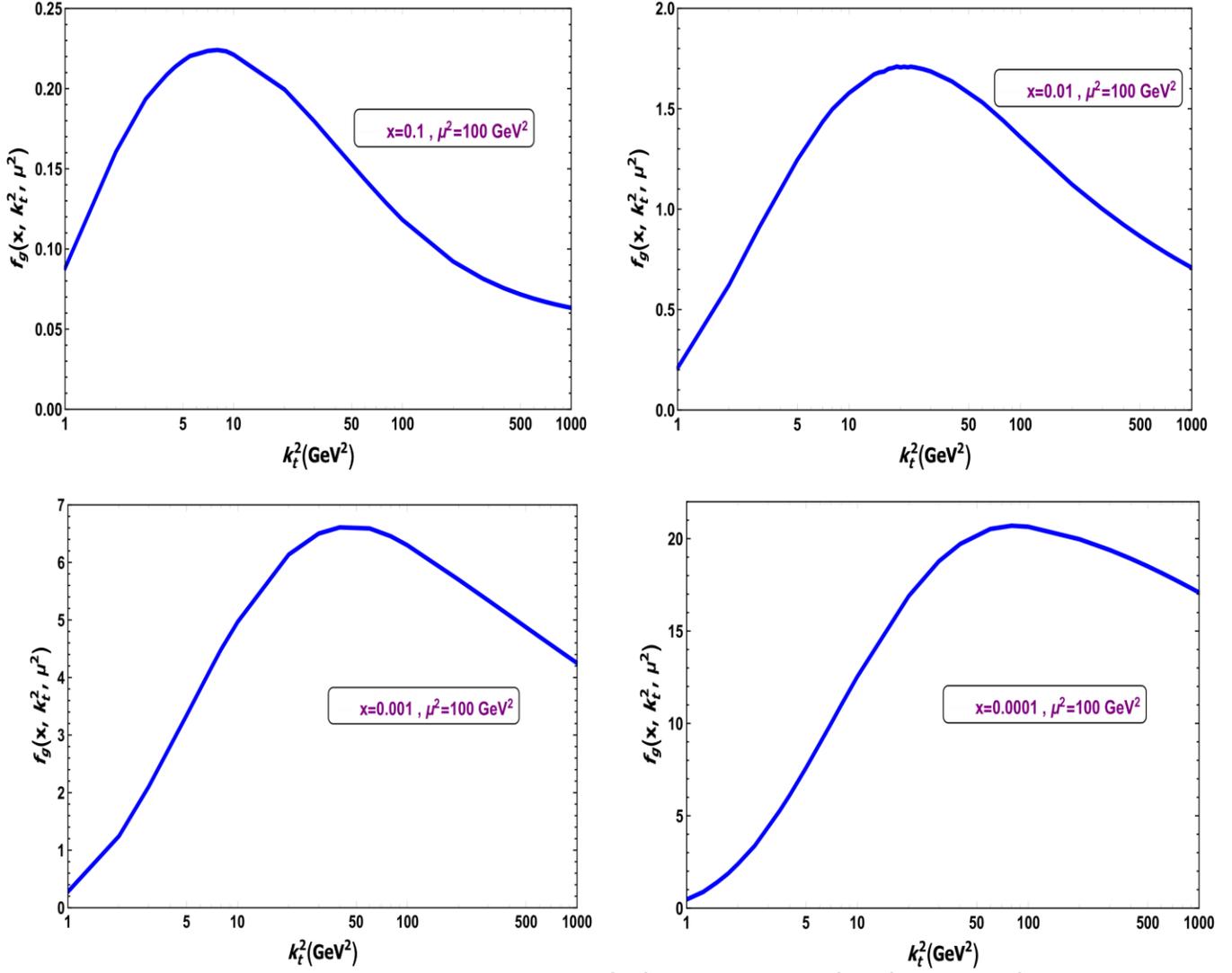

Figure 3: The unintegrated gluon distribution function, $f_g(x, k_t^2, \mu^2)$, with respect to $k_t^2$ at $\mu^2 = 100$ GeV$^2$, for various values of $x$.

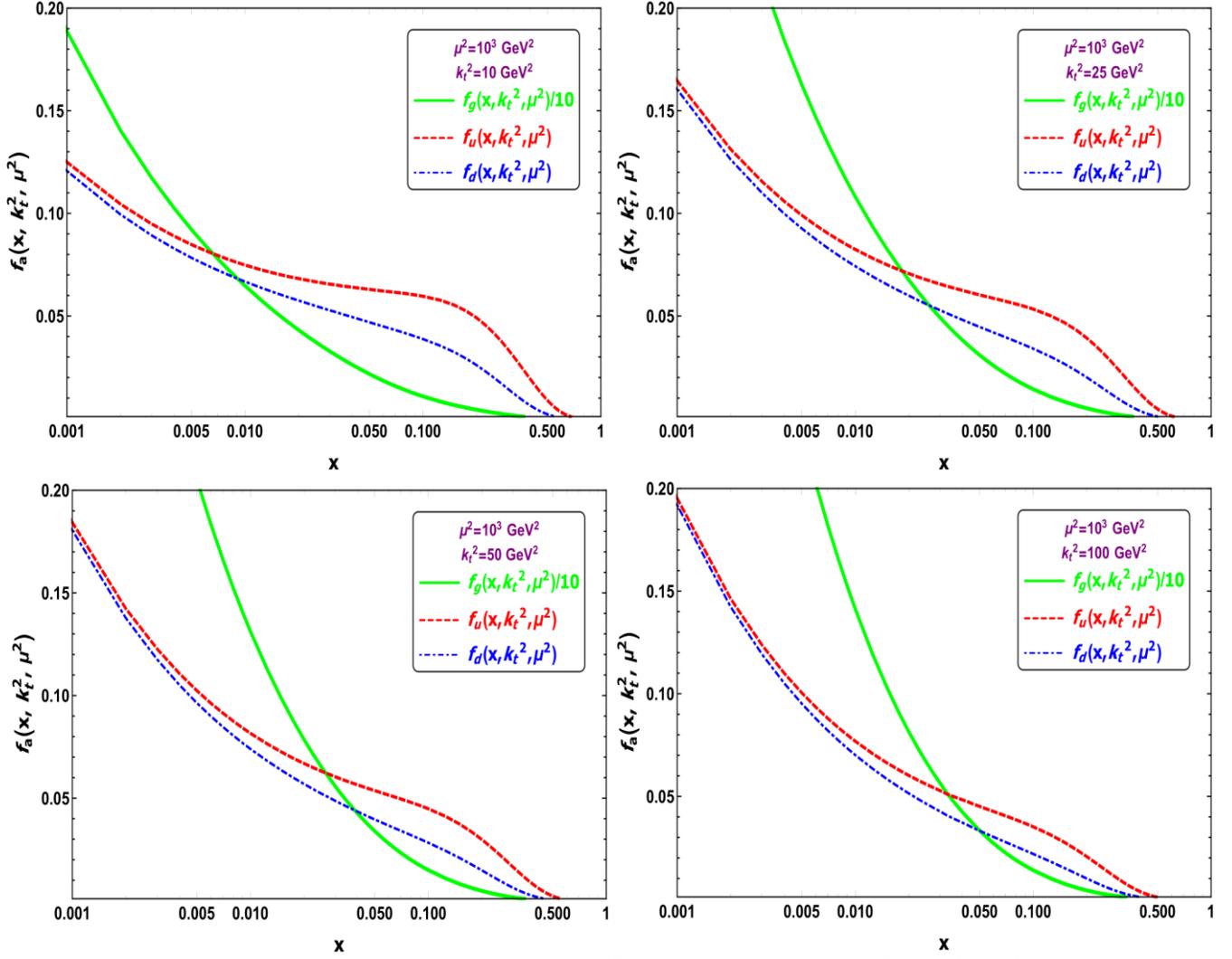

Figure 4: The unintegrated parton distribution function, $f_a(x, k_t^2, \mu^2)$, with respect to $x$ at $\mu^2 = 10^3$ GeV$^2$, for different values of $k_t^2$. The solid line corresponds to the unintegrated gluon distribution, $f_g(x, k_t^2, \mu^2)$, divided on 10, dashed line to the unintegrated up quark distribution, $f_u(x, k_t^2, \mu^2)$, and the dot-dashed line to the unintegrated down quark distribution, $f_d(x, k_t^2, \mu^2)$.

## 3- The total amplitude of the LO subprocesses

There are two LO subprocesses, Compton and annihilation scattering, which lead to the single direct photon in the $p\bar{p}$ collision. Their Feynman diagrams are illustrated in table 1:

Table 1: Two LO subprocesses leading to the single direct photon in $p + \bar{p} \to \gamma + X$

| | |
|---|---|
| $q(k_1) + g(k_2) \to \gamma(p^\gamma) + q(p^c)$ | |
| $q(k_1) + \bar{q}(k_2) \to \gamma(p^\gamma) + g(p^c)$ | |

The amplitudes of Compton collision are illustrated in tables 2:

Table 2: The amplitude of each Compton collision related to corresponded Feynman diagram. The first amplitude corresponds to the right Feynman diagram, the second to the left.

| | |
|---|---|
| $\mathcal{M}_1(q(k_1) + g(k_2) \to \gamma(p^\gamma) + q(p^c))$ | $e_q g\, T^a\, \bar{u}(p^c)\, \gamma^\mu \varepsilon_\mu(k_2)\, \dfrac{1}{(\not{k}_1 - \not{p}^\gamma) - m}\, \varepsilon_\nu^*(p^\gamma)\, \gamma^\nu u(k_1)$ |
| $\mathcal{M}_2(q(k_1) + g(k_2) \to \gamma(p^\gamma) + q(p^c))$ | $e_q g\, T^b\, \bar{u}(p^c)\gamma^\nu \varepsilon_\nu(p^\gamma)\, \dfrac{1}{(\not{k}_1 + \not{k}_2) - m}\, \varepsilon_\mu^*(k_2)\, \gamma^\mu u(k_1)$ |

And the amplitudes of annihilation collision are presented in tables 3:

Table 3: The amplitude of each annihilation collision related to corresponded Feynman diagram. The first amplitude corresponds to the left Feynman diagram, the second to the right.

| | |
|---|---|
| $\mathcal{M}_1(q(k_1) + \bar{q}(k_2) \to \Upsilon(p^\gamma) + g(p^c))$ | $e_q g\, T^a\, \bar{v}(k_2)\gamma^\nu \varepsilon_\nu(p^c)\, \dfrac{1}{(\not{k}_1 - \not{p}^\gamma) - m}\, \varepsilon_\mu^*(p^\gamma)\, \gamma^\mu u(k_1)$ |
| $\mathcal{M}_2(q(k_1) + \bar{q}(k_2) \to \Upsilon(p^\gamma) + g(p^c))$ | $e_q g\, T^b\, \bar{v}(k_2)\gamma^\nu \varepsilon_\nu(p^\gamma)\, \dfrac{1}{(\not{k}_1 - \not{p}^c) - m}\, \varepsilon_\mu^*(p^c)\, \gamma^\mu u(k_1)$ |

Where $g^2 = 4\pi\alpha_s$, $T^a = \lambda^a/2$ and $e_q$ is the fraction charge of the quarks, for example $e_d = -1/3$.

Incoming on-shell partons do not depend on transverse momentum, so their momentum is:

$$k = xP. \qquad (25)$$

Where $P$ is the momentum of proton or antiproton and $x$ the longitudinal momentum fraction. Because of high energy collision, the mass of proton and antiproton is neglected, $P^2 = m_p^2 = 0$. So, the mass of on-shell partons is considered zero in calculations. The total amplitude of each LO subprocess by on-shell incoming partons is given in table 4:

Table 4: The total amplitude of the Compton and annihilation collision for the on-shell incoming partons.

| | |
|---|---|
| $\|\bar{\mathcal{M}}\|^2(q(k_1) + g(k_2) \rightarrow \gamma(p^\gamma) + q(p^c)))$ | $\dfrac{-16\pi^2\, \alpha_{em}\alpha_s e_q^2}{3}\left(\dfrac{\hat{t}}{\hat{s}} + \dfrac{\hat{s}}{\hat{t}}\right)$ |
| $\|\bar{\mathcal{M}}\|^2(q(k_1) + \bar{q}(k_2) \rightarrow \Upsilon(p^\gamma) + g(p^c))$ | $\dfrac{128\pi^2\, \alpha_{em}\alpha_s e_q^2}{9}\left(\dfrac{\hat{t}}{\hat{u}} + \dfrac{\hat{u}}{\hat{t}}\right)$ |

Off-shell incoming partons depend on the transverse momentum as the following relation :

$$k = xP + k_t. \qquad (26)$$

$k_T$ is the transverse momentum of the parton. Since the off-shell partons depend on transverse momentum, which leads the partons to have virtual mass $m^2 = k_t^2 = -\mathbf{k}_T^2$, the total amplitude of LO off-shell subprocesses takes the mass terms. The total amplitude of each off-shell collision is presented in table 5:

Table 5: The total amplitude of the Compton and annihilation collision for the off-shell incoming partons.

| | |
|---|---|
| $\|\bar{\mathcal{M}}\|^2(q(k_1) + g(k_2) \rightarrow \gamma(p^\gamma) + q(p^c)))$ | $\dfrac{16\pi^2\, \alpha_{em}\alpha_s e_q^2}{3(\hat{t}-m^2)^2(\hat{s}-m^2)^2}\, F_{qg}(\mathbf{k}_{1T}^2, \mathbf{k}_{2T}^2)$ |
| $\|\bar{\mathcal{M}}\|^2(q(k_1) + \bar{q}(k_2) \rightarrow \Upsilon(p^\gamma) + g(p^c))$ | $\dfrac{-128\pi^2\, \alpha_{em}\alpha_s e_q^2}{9(\hat{t}-m^2)^2(\hat{u}-m^2)^2}\, F_{q\bar{q}}(\mathbf{k}_{1T}^2, \mathbf{k}_{2T}^2)$ |

Where $F_{qg}(\mathbf{k}_{1T}^2, \mathbf{k}_{2T}^2)$ and $F_{q\bar{q}}(\mathbf{k}_{1T}^2, \mathbf{k}_{2T}^2)$ are:

$$F_{qg}(\mathbf{k}_{1T}^2, \mathbf{k}_{2T}^2) = 6m^8 - (2\mathbf{k}_{2T}^4 + 2(\hat{s}+\hat{t})\mathbf{k}_{2T}^2 + 3\hat{s}^2 + 3\hat{t}^2 + 14\hat{s}\hat{t})m^4$$
$$+ (2(\hat{s}+\hat{t})\mathbf{k}_{2T}^4 + 8\hat{s}\hat{t}\mathbf{k}_{2T}^2 + \hat{s}^3 + \hat{t}^3 + 7\hat{s}\hat{t}^2 + 7\hat{t}\hat{s}^2)m^2 - \hat{s}\hat{t}(2\mathbf{k}_{2T}^4 + 2(\hat{s}+\hat{t})\mathbf{k}_{2T}^2 + \hat{s}^2 + \hat{t}^2). \qquad (27)$$

And

$$F_{q\bar{q}}(\mathbf{k}_{1T}^2, \mathbf{k}_{2T}^2) = 6m^8 - (3\hat{t}^2 + 3\hat{u}^2 + 14\hat{u}\hat{t})m^4 + (\hat{u}^3 + \hat{t}^3 + 7\hat{u}\hat{t}^2 + 7\hat{t}\hat{u}^2)m^2 - \hat{u}\hat{t}(\hat{u}^2 + \hat{t}^2). \qquad (28)$$

The items $\hat{t}$, $\hat{u}$, and $\hat{s}$ are the Mandelstam variables with the following equations:

$$\hat{s} = (k_1 + k_2)^2, \quad \hat{t} = (k_1 - p^\gamma)^2, \quad \hat{u} = (k_1 - p^c)^2. \qquad (29)$$

The term $m$ is the mass of the quark in Equ. (27) and (28). The total amplitudes of two LO collisions in table 5 and Equ. (27) and (28) are the same as the amplitudes in [32]. To obtain the total amplitude for collisions of $g + q \rightarrow \gamma + q$ and $\bar{q} + q \rightarrow g + \gamma$, it is enough to change $\hat{u} \leftrightarrow \hat{t}$ in table 4 & 5. All of the amplitudes and total amplitudes in tables 2-5 are obtained and calculated by FeynCalc [83, 84]

# 4- The cross section of the single direct photon

In this section, the kinematics and equations to calculate the cross section of the single direct photon in the $p\bar{p}$ collision are presented. They are given for both the on-shell and off-shell incoming partons. The incoming particles are considered two high energy proton and antiproton colliding each other at the center of mass frame. Then, their mass is neglected, and they have four-momentum:

$$P_1 = \frac{\sqrt{s}}{2}(1,0,0,1) \text{ and } P_2 = \frac{\sqrt{s}}{2}(1,0,0,-1), \tag{30}$$

## 4-1 Kinematics of the particles

The momentums of the incoming particles are:

$$k_1 = x_1 P_1 + k_{1t} , \quad k_2 = x_2 P_2 + k_{2t} \text{ (for off-shell partons)} \tag{31}$$

$$k_1 = x_1 P_1 , \quad k_2 = x_2 P_2 \text{ (for on-shell partons)} \tag{32}$$

And for the outgoing particles:

$$p^\gamma = \alpha_1 P_1 + \beta_1 P_2 + p_T^\gamma \qquad p^c = \alpha_2 P_1 + \beta_2 P_2 + p_T^c \tag{33}$$

By conversation laws, the following relation is obtained:

$$x_1 = \alpha_1 + \alpha_2 \;\&\; x_2 = \beta_1 + \beta_2 \;\&\; p_T^\gamma + p_T^c = k_{1t} + k_{2t} \tag{34}$$

Where the Sudakov variables are:

$$\alpha_1 = \frac{E_T^\gamma}{\sqrt{s}} \exp(y^\gamma) \qquad \alpha_2 = \frac{m_T^c}{\sqrt{s}} \exp(y^c) . \tag{35}$$

$$\beta_1 = \frac{E_T^\gamma}{\sqrt{s}} \exp(-y^\gamma) \quad \beta_2 = \frac{m_T^c}{\sqrt{s}} \exp(-y^c). \tag{36}$$

The items of $m_T$ (transverse mass) and $y$ (rapidity) have the following relations:

$$m_T = \sqrt{m^2 + p_T^2} \;\&\; y = \frac{1}{2}\ln\left(\frac{E + p_z}{E - p_z}\right). \tag{37}$$

In the off-shell state, $k_1^2 = k_{1t}^2 = -\boldsymbol{k}_{1T}^2 \neq 0$ and $k_2^2 = k_{2t}^2 = -\boldsymbol{k}_{2T}^2 \neq 0$, so the two incoming partons are massive. When the mass of a particle is zero, the rapidity and pseudorapidity would be equal:

$$y = \eta = -\ln\tan\left(\frac{\theta}{2}\right). \tag{38}$$

Where $\theta$ is the polar angle with respect to the proton beam. The four-momentum of each particle, whether be on-shell or off-shell, is:

$$p^\mu = (E, p_x, p_y, p_z) = (m_T \cosh y , p_T \sin\varphi , p_T \cos\varphi , m_T \sinh y). \tag{39}$$

## 4-2 The cross section of off-shell incoming partons

For two LO subprocesses, the cross section of the single direct photon by two off-shell incoming partons in $p\bar{p}$ collisions is:

$$d\sigma(p+\bar{p} \to \gamma + X) = \sum_{ab} \int \frac{dx_1 d\mathbf{k}_{1T}^2}{x_1} \frac{f_a(x_1, \mathbf{k}_{1T}^2, \mu^2)}{\mathbf{k}_{1T}^2} \times \int \frac{dx_2 d\mathbf{k}_{2T}^2}{x_2} \frac{f_b(x_2, \mathbf{k}_{2T}^2, \mu^2)}{\mathbf{k}_{2T}^2} \times d\sigma(ab \to \gamma c) \quad (40)$$

Where particle $a$ is the parton emitted from the proton, particle $b$ is the parton from the antiproton, and particle c is the outgoing particle. The term $d\sigma(ab \to \gamma c)$ is the differential cross section of subprocess corresponded to partons $a$ and $b$, can be written as:

$$d\sigma = \frac{d\cos(\theta^*)}{32\pi\, x_1 x_2 s} |\bar{\mathcal{M}}|^2 , \quad (41)$$

$\theta^*$ is the angle between the outgoing particles and proton beam in the center of mass frame. The differential cross section should be expressed in the term of measurable variables such as rapidity and transverse momentum of the outgoing particle, then by using the determinant of the Jacobian matrix [85]:

$$dx_1 dx_2 d\cos(\theta^*) = \frac{4 E_T^\gamma dE_T^\gamma dy^\gamma dy^c}{s} , \quad (42)$$

By Equ. (41) and (42) the cross section of the single photon with off-shell partons will be:

$$\sigma(p+\bar{p} \to \gamma + X)$$
$$= \sum_{ab} \int \frac{E_T^\gamma}{8\pi(x_1 x_2 s)^2} |\bar{\mathcal{M}}|^2(ab \to \gamma c)$$
$$\times \frac{f_a(x_1, \mathbf{k}_{1T}^2, \mu^2)}{\mathbf{k}_{1T}^2} \frac{f_b(x_2, \mathbf{k}_{2T}^2, \mu^2)}{\mathbf{k}_{2T}^2} d\mathbf{k}_{1T}^2 d\mathbf{k}_{2T}^2 dE_T^\gamma dy^c d\eta^\gamma \frac{d\varphi_1}{2\pi} \frac{d\varphi_2}{2\pi} \frac{d\varphi^\gamma}{2\pi}. \quad (43)$$

## 4-3 The cross section of on-shell incoming partons

In the case of on-shell incoming partons, it is enough to take $k_t$ equal to zero. It causes the second and third components of four-momentum of the incoming partons to become zero which removes the dependence on $\varphi_1$ and $\varphi_2$ (look at Equ. (39) and take $k_t = 0$). In the Equ. (40) instead of using Equ. (9) to find the number of partons contributing to subprocesses, one must use Equ. (8). Therefore, the cross section of the single photon with on-shell partons will be:

$$\sigma(p+\bar{p} \to \gamma + X)$$
$$= \sum_{ab} \int \frac{E_T^\gamma}{8\pi(x_1 x_2 s)^2} |\bar{\mathcal{M}}|^2(ab \to \gamma c) \times f_a(x_1, \mu^2) f_b(x_2, \mu^2)\, dE_T^\gamma dy^c d\eta^\gamma \frac{d\varphi^\gamma}{2\pi}. \quad (44)$$

Where $f_a(x_1, \mu^2) = xg(x_1, \mu^2)$ or $xq(x_1, \mu^2)$. It is visible that the on-shell cross section does not depend on $k_t$.

## 5- Numerical results and discussion

In this section, some explanations about theoretical calculations are given then the results will be presented and discussed. First, it is used the LO formula for $\alpha_s$ with $n_f = 5$ and $\Lambda = 0.226$ GeV in all equations. To consider theoretical uncertainties in the calculation of the cross section, three different factorisation and renormalization scales are used, the first; $\mu_F = \mu_R = \mu = E_T^\lambda$, the second, $\mu_F = \mu_R = \mu = 0.5\, E_T^\lambda$, and the last, $\mu_F = \mu_R = \mu = 2\, E_T^\lambda$. All quarks are considered in calculations except the quark "t". The mass of quarks is assumed to be zero which does not make a significant difference. To calculate the double differential cross section, one must use:

$$\frac{d^2\sigma}{dE_T^\gamma d\eta^\gamma} = \frac{1}{\Delta \eta^\gamma} \int_{-\eta^\gamma}^{\eta^\gamma} \frac{d\sigma}{dE_T^\gamma}\, d\eta^\gamma. \qquad (45)$$

The effect of the fragmentation processes was not considered in the calculations of this paper. Moreover, the theoretical results are compared with the experimental data obtained by some imposed constraints to reduce the effect of the fragmentation. All calculations, such as calculating of Equ. (43) and (44), total amplitudes, are computed by using Mathematica [86].

Theoretical results are compared with three experimental data [4, 7, 10]. In the first one, DØ [4], the energy of the collision is $\sqrt{s} = 630$ GeV, with central and forward pseudo-rapidity $|\eta^\gamma| < 0.9$ and $1.6 < |\eta^\gamma| < 2.5$, respectively. In the second one, DØ [7], the energy of the collision is $\sqrt{s} = 1960$ GeV, with central pseudo-rapidity $|\eta^\gamma| < 0.9$. In the last one, CDF [10], the energy of the collision is $\sqrt{s} = 1960$ GeV, with central pseudo-rapidity $|\eta^\gamma| < 1$. In all mentioned experiments, the data for double differential cross sections with respect to the transverse momentum of the photon, $E_T^\lambda$, is presented. To better understanding and comparing what goes on in the figures, I give my results numerically in tables 6-8 in the appendix. The cross section of the single photon is calculated in both collinear (on-shell) and $k_t$-factorisation frame, by the Equ. (43) and (44), respectively. The ratio of the off-shell to the on-shell cross section is given for three different experimental conditions in figure 6. The LO subprocesses leading to the single photon are analyzed in figure 5 to see the dominant collision at the corresponded region of $E_T^\lambda$. The ratio of the experimental data to the theoretical cross section is illustrated in figure 8, and the impact of the theoretical uncertainties on the value of the cross section is investigated in figure 7.

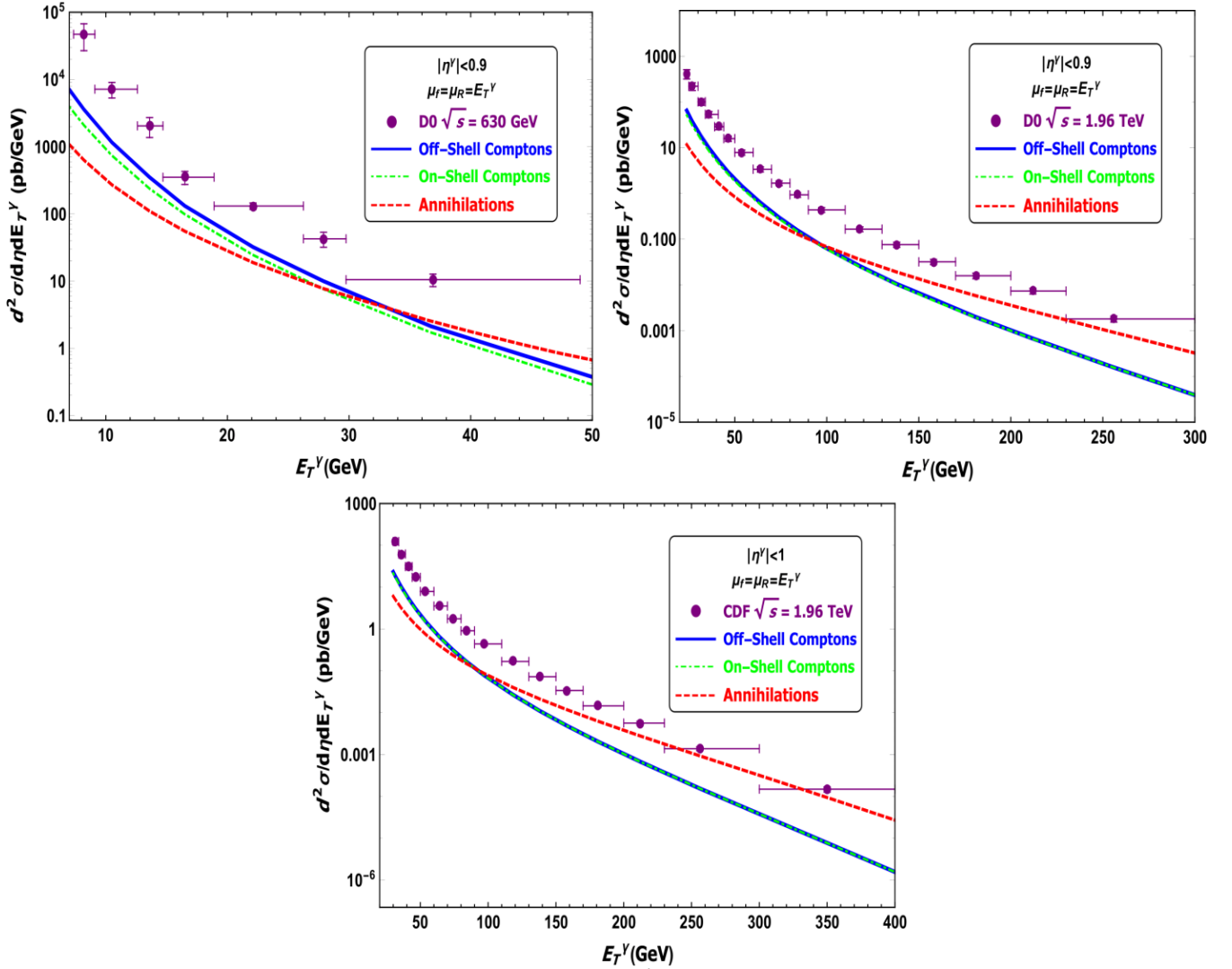

Figure 5: The double differential cross section with respect to $E_T^\lambda$ for Compton and annihilation subprocesses at the three different experiments. The solid line corresponds to the off-shell comptons, dot-dashed to the on-shell comptons, and dashed to the annihilations.

In the case of $p\bar{p}$ collisions, because of the symmetry between an antiquark within an antiproton and the same flavor quark within a proton, the Compton collisions are not always the dominant subprocesses while in $pp$ collisions they are always dominant. It is seen in figure 5, the annihilation collisions become more dominant than the comptons as $E_T^\lambda$ increases. The cross sections in figure 5 are calculated by Equ. (44) and (43) for the off-shell and on-shell collisions, respectively. The differences in the calculations of the two mentioned cross sections come from different total amplitudes and parton distribution functions. According to figure 1 and 4, it is seen that as $x$ increases, the distribution function of both quarks up and down becomes larger than the gluon especially when $k_t$ is small. On the other hand, in Equ. (34-36) one can see that $\sim x_T = E_T^\lambda/\sqrt{s}$, consequently increasing in the value of $E_T^\lambda$, which is equivalent to the increase in the value of $x$, leads to the dominating of the distribution function for both quarks up and down than gluon. Then the Compton processes are dominant at the modest and small $E_T^\lambda$ while at large $E_T^\lambda$, it is the annihilations that control the LO subprocesses.

In figure 6, the double differential cross section of LO subprocesses leading to the single direct photon for both the off-shell and on-shell incoming partons (left plots), in addition, their ratio (right plots) are illustrated. The plots are presented for DØ with $\sqrt{s} = 630$ GeV, DØ with $\sqrt{s} = 1.96$ TeV, and CDF with $\sqrt{s} = 1.96$ TeV. It is seen see that theoretical result agree well with the experimental data for each of the three experiments especially at large $E_T^\lambda$. As it is seen, the ratio of the off-shell cross section to the on-shell decreases as $x_T$ increases. The total value of the ratio For two collisions with higher collision energy than DØ with $\sqrt{s} = 630$ GeV is declined as the energy of collision increases. Recall that $E_T^\lambda$ and $x_T$ and $x$ are equivalent to each other by Equ. (34-36). Then the differences between off-shell and on-shell incoming partons appear at small $x$, while there is almost no difference between off-shell and on-shell particles at large $x$.

The calculations of the cross section of single direct photon depend on two different parameters $\mu_F$ and $\mu_R$. The first one is related to factorisation scale, $\mu_F$, in Equ. (10) and the second to the strong coupling constant $\alpha_s(\mu_R^2)$. Since there is no certain value for them, they are assumed approximately and are theoretical errors. In figure 7, the theoretical errors in calculations of the cross section are considered for both off-shell and on-shell states. We see that the value of cross sections increases as we choose the smaller value for $\mu$. It is true for both off-shell and on-shell cross sections. Changing $\mu_F = \mu_R = \mu = E_T^\lambda$ to $\mu_F = \mu_R = \mu = 0.5\ E_T^\lambda$ or $\mu_F = \mu_R = \mu = 2\ E_T^\lambda$ modifies the value of cross sections from 10% for lower $E_T^\lambda$ to 30% for high $E_T^\lambda$ which is almost true for both the off-shell and on-shell in each different collision DØ and CDF.

To get a better understanding of the agreement and accuracy of the theoretical calculated cross section to the experimental data, the ratio of data to theory is calculated in figure 8. As it is seen in all plots, the ratio will decrease as $E_T^\lambda$ increases. It indicates, at high $E_T^\lambda$, the portion of LO collisions in the total value of the cross section will increase than other subprocesses, such as NLO, NNLO, and fragmentation processes. Remember that the LO subprocesses in theoretical calculations are only considered. Moreover, at low $E_T^\lambda$, the ratio is large, so in this region, the impact of fragmentation processes may be considerable. Also, we see that the off-shell cross sections have a smaller ratio, especially for the low value of $E_T^\lambda$, than on-shells in each plot of figure 8. Then the off-shell cross sections agree better than on-shells with data. But it is seen, at the higher $E_T^\lambda$, that both off-shell and on-shell cross sections approach equally to the experimental value of the cross section. Notice that there are other reasons leading to inequality of theoretical and experimental data, such as uncertainty, which was 68% confidence level, in the value of integrated parton distribution functions.

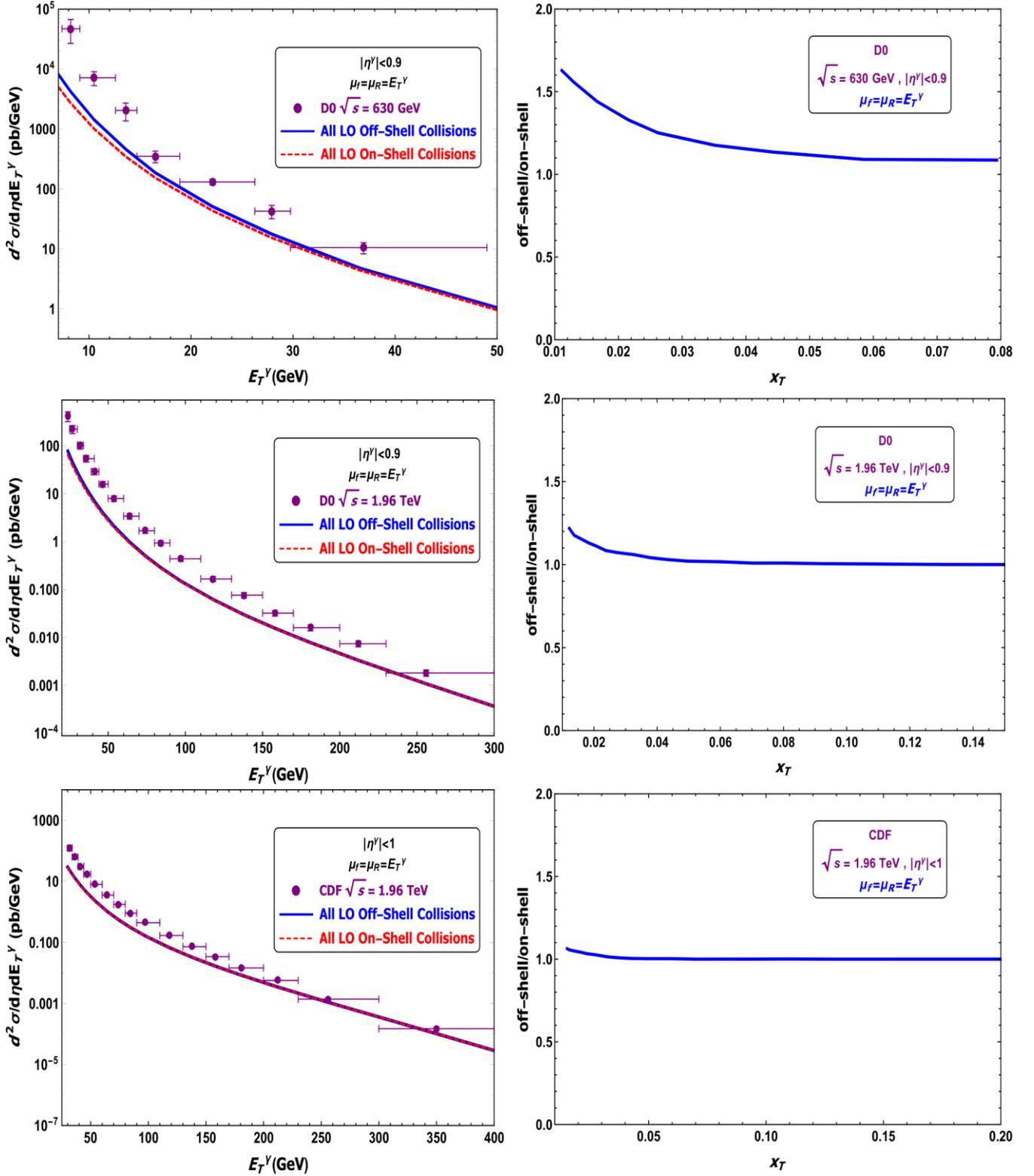

Figure 6: In the left, the double differential cross section of all LO direct collisions leading to the single photon. The solid lines correspond to off-shell collisions and dashed lines to the on-shell collision. In the right, the ratio of all LO off-shell cross section to all LO on-shell cross section with respect to $x_T$.

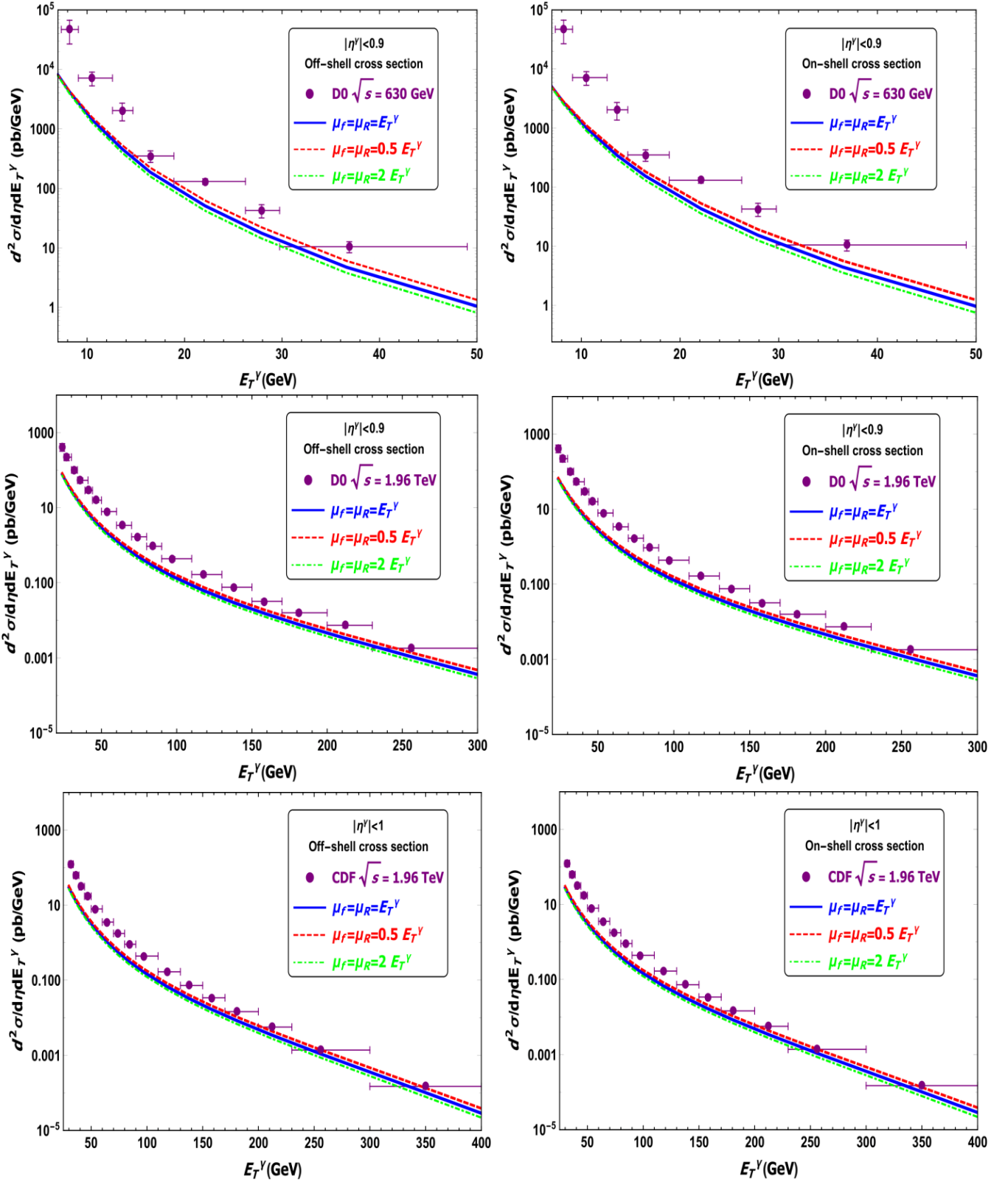

Figure 7: The double differential cross section of the single direct photon with various uncertainties (theoretical error) for both the off-shell (left plots) and on-shell (right plots) LO collisions. The solid blue lines correspond to $\mu_F = \mu_R = \mu = E_T^\lambda$, the dashed red lines to $\mu_F = \mu_R = \mu = 0.5\, E_T^\lambda$, and dot-dashed green lines to $\mu_F = \mu_R = \mu = 2\, E_T^\lambda$.

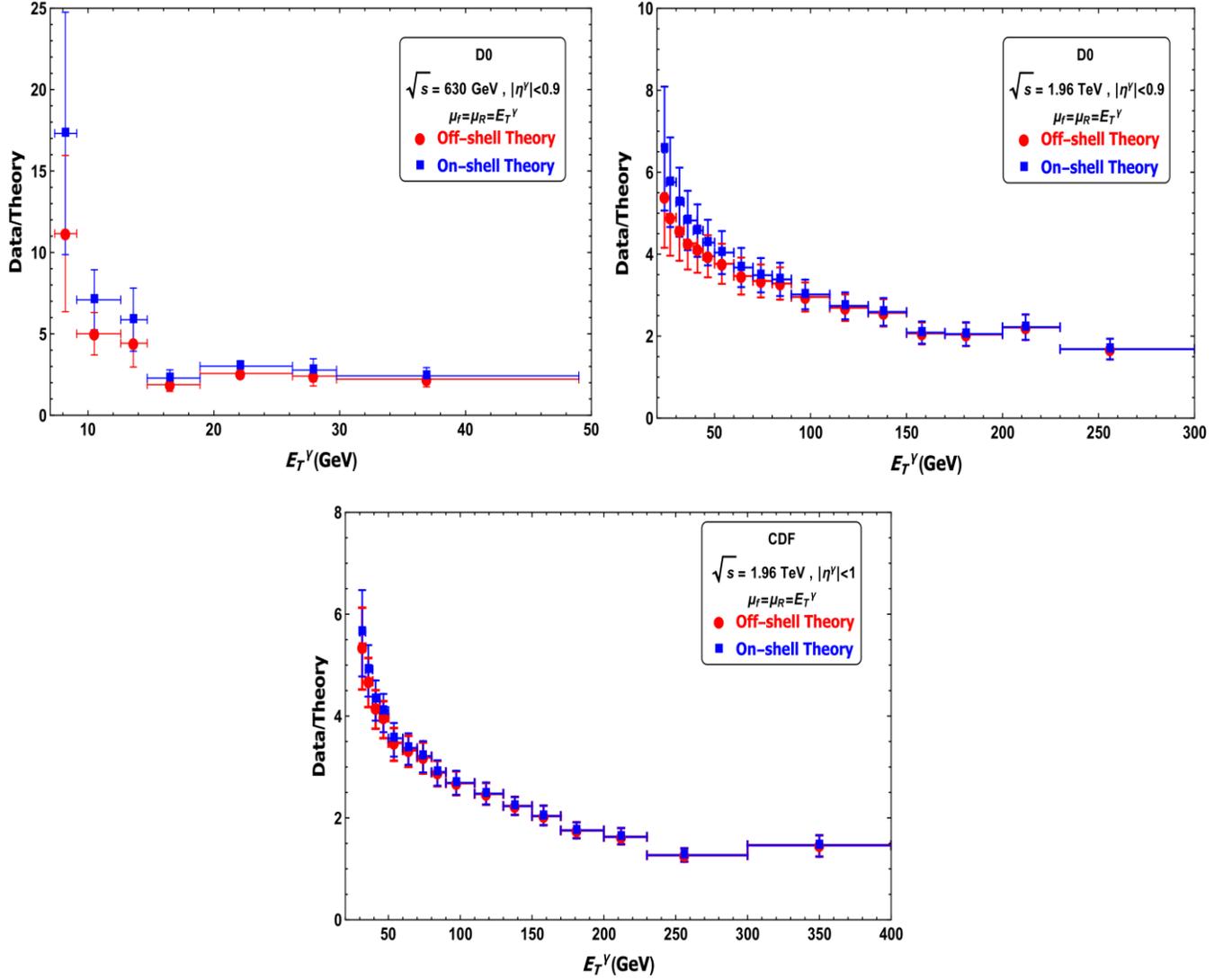

Figure 8: the ratio of the measured experimental cross section of single isolated photon to the calculated theoretical cross section in three experiments of DØ with $\sqrt{s} = 630$ GeV, DØ $\sqrt{s} = 19.6$ TeV, and CDF $\sqrt{s} = 1.96$ TeV. The ratio is for both off-shell (filled circle) and on-shell (filled square). The vertical lines indicate the uncorrelated uncertainty in DØ with $\sqrt{s} = 630$ GeV and the systematic uncertainties for two other plots.

## 6- Summary and Conclusions

The double differential cross section of the single direct photon in the $p\bar{p}$ collisions was calculated by using the LO MRW method and MMHT 2014 LO PDFs for the off-shell and on-shell incoming partons, respectively. The theoretical results were compared with the corresponded experimental data and found out that they agree well with data in the case of both the off-shell and on-shell cross sections. Moreover, it was established that the Compton collisions are more dominant than annihilations at the low and modest $E_T^\lambda$ while at larger $E_T^\lambda$ the annihilations are. Although the MRW method is valid for all ranges of $x$, it was determined that the differences between being on-shell or off-shell would appear at low $x$ ( because of $x \sim x_T = E_T^\gamma/\sqrt{s}$) by the decline in the ratio of the off-shell to the on-shell cross section, approach to one, as $x_T$ ($x$) increases. It was seen that the off-shell cross sections were closer to experimental data especially for low $E_T^\gamma$. The theoretical uncertainties changed the cross section 10% to 30% and all approach each other at low $E_T^\gamma$. Both of off-shell and on-shell cross sections were closer to data at large $E_T^\gamma$ than small $E_T^\gamma$. It indicates that the LO subprocesses are very important at large $E_T^\gamma$ while fragmentation and the higher-order subprocesses (NLO, NNLO) have a considerable impact at low $E_T^\gamma$ or $x_T$. Finally, it is concluded that although taking off-shell partons in the calculations of subprocesses, especially at small $x_T$, are more accurate than on-shells, one can trust results of on-shell calculations for large $x_T$.

# Appendix

In this section, to give a more accurate understanding of the exact value of the results that are plotted in the figures, I present the numerical results of my calculations in table 6-8 for the three considered experiments. In the 9$^{th}$, 10$^{th}$, and 11$^{th}$ columns, both of the on-shell and off-shell theoretical cross section are calculated by $\mu_F = \mu_R = E_T^\gamma$.

Table 6: The numerical results of the double differential cross section calculated with three different theoretical uncertainties for both the off-shell and on-shell state. The experimental data is based on [4]. Other information about the calculations is given at the beginning of section 5.

| | **All LO collisions** DØ ($\sqrt{s} = 630$ GeV, $|\eta^\gamma| < 0.9$) | | | | | | | | | |
|---|---|---|---|---|---|---|---|---|---|---|
| $E_T^\gamma$ (GeV) | $\frac{d^2\sigma}{dE_T^\gamma d\eta^\gamma}$ ($\frac{pb}{GeV}$) (on-shell) | | | $\frac{d^2\sigma}{dE_T^\gamma d\eta^\gamma}$ ($\frac{pb}{GeV}$) (off-shell) | | | Experimental data | Data on off-shell Theory for $\mu_F = \mu_R = E_T^\lambda$ | Data on on-shell Theory for $\mu_F = \mu_R = E_T^\lambda$ | Off-shell on on-shell for $\mu_F = \mu_R = E_T^\lambda$ |
| | $\mu_F = \mu_R = 0.5 E_T^\lambda$ | $\mu_F = \mu_R = E_T^\lambda$ | $\mu_F = \mu_R = 2 E_T^\lambda$ | $\mu_F = \mu_R = 0.5 E_T^\lambda$ | $\mu_F = \mu_R = E_T^\lambda$ | $\mu_F = \mu_R = 2 E_T^\lambda$ | | | | |
| 8.2 | 2505 | 2715 | 2508 | 4367 | 4212 | 3905 | 47000 | 11.15 | 17.3 | 1.56 |
| 10.5 | 1110 | 1010 | 898 | 1576 | 1430 | 1275 | 7160 | 5 | 7.09 | 1.44 |
| 13.6 | 401 | 347.5 | 299.5 | 532.8 | 461 | 397.9 | 2040 | 4.42 | 5.87 | 1.32 |
| 16.5 | 182.5 | 153.7 | 129.9 | 221.8 | 187.5 | 157.96 | 351 | 1.88 | 2.28 | 1.25 |
| 22.1 | 53.9 | 43.4 | 35.86 | 62.68 | 51.02 | 42.19 | 131 | 2.57 | 3.02 | 1.18 |
| 27.9 | 19.23 | 15.34 | 12.49 | 22.2 | 17.62 | 14.43 | 42.6 | 2.4 | 2.77 | 1.13 |
| 36.9 | 5.65 | 4.34 | 3.55 | 6.05 | 4.74 | 3.8 | 10.5 | 2.22 | 2.42 | 1.09 |

Table 7: The numerical results of the double differential cross section calculated with three different theoretical uncertainties for both the off-shell and on-shell state. The experimental data is based on [7]. Other information about the calculations is given at the beginning of section 5.

| | **All LO collisions** | | | | | | | | | |
|---|---|---|---|---|---|---|---|---|---|---|
| | **DØ ($\sqrt{s} = 1.96$ TeV, $|\eta^\gamma| < 0.9$)** | | | | | | | | | |
| $E_T^\gamma$ (GeV) | $\frac{d^2\sigma}{dE_T^\gamma d\eta^\gamma}$ ($\frac{\text{pb}}{\text{GeV}}$) (on-shell) | | | $\frac{d^2\sigma}{dE_T^\gamma d\eta^\gamma}$ ($\frac{\text{pb}}{\text{GeV}}$) (off-shell) | | | Experimental data | Data on off-shell Theory for $\mu_F = \mu_R = E_T^\lambda$ | Data on on-shell Theory for $\mu_F = \mu_R = E_T^\lambda$ | Off-shell on on-shell for $\mu_F = \mu_R = E_T^\lambda$ |
| | $\mu_F = \mu_R = 0.5 E_T^\lambda$ | $\mu_F = \mu_R = E_T^\lambda$ | $\mu_F = \mu_R = 2 E_T^\lambda$ | $\mu_F = \mu_R = 0.5 E_T^\lambda$ | $\mu_F = \mu_R = E_T^\lambda$ | $\mu_F = \mu_R = 2 E_T^\lambda$ | | | | |
| 23.9 | 67.44 | 62.95 | 57.96 | 82.16 | 76.69 | 70.62 | 414 | 5.4 | 6.57 | 1.21 |
| 26.9 | 41.85 | 38.4 | 34.92 | 49.23 | 45.17 | 41.08 | 221 | 4.89 | 5.56 | 1.18 |
| 31.7 | 21.34 | 19.17 | 17.17 | 24.59 | 22.09 | 19.78 | 101 | 4.57 | 5.27 | 1.15 |
| 36 | 12.57 | 11.13 | 9.86 | 14.21 | 12.59 | 11.15 | 53.7 | 4.27 | 4.82 | 1.13 |
| 41.1 | 7.20 | 6.29 | 5.52 | 7.99 | 6.99 | 6.13 | 28.8 | 4.12 | 4.57 | 1.11 |
| 46.5 | 4.27 | 3.69 | 3.21 | 4.63 | 4.00 | 3.48 | 15.8 | 3.95 | 4.28 | 1.08 |
| 53.8 | 2.29 | 1.96 | 1.69 | 2.45 | 2.10 | 1.81 | 7.9 | 3.76 | 4.04 | 1.07 |
| 63.9 | 1.09 | 0.923 | 0.788 | 1.16 | 0.978 | 0.835 | 3.39 | 3.47 | 3.67 | 1.06 |
| 74.1 | 0.57 | 0.48 | 0.41 | 0.6 | 0.50 | 0.426 | 1.68 | 3.34 | 3.49 | 1.04 |
| 84.1 | 0.33 | 0.276 | 0.232 | 0.342 | 0.284 | 0.240 | 0.934 | 3.28 | 3.39 | 1.03 |
| 97.2 | 0.175 | 0.145 | 0.122 | 0.179 | 0.148 | 0.124 | 0.438 | 2.96 | 3.02 | 1.02 |
| 118 | 0.074 | 0.061 | 0.050 | 0.075 | 0.062 | 0.051 | 0.166 | 2.69 | 2.74 | 1.01 |
| 138 | 0.036 | 0.0294 | 0.0243 | 0.037 | 0.030 | 0.0245 | 0.0761 | 2.57 | 2.59 | 1.009 |
| 158 | 0.0190 | 0.0153 | 0.0125 | 0.0192 | 0.0154 | 0.0127 | 0.032 | 2.07 | 2.09 | 1.009 |
| 181 | 0.0096 | 0.0077 | 0.0063 | 0.0097 | 0.0077 | 0.0064 | 0.0159 | 2.04 | 2.05 | 1.005 |
| 212 | 0.0041 | 0.0033 | 0.0027 | 0.0042 | 0.0033 | 0.0027 | 0.0074 | 2.21 | 2.22 | 1.003 |
| 256 | 0.0013 | 0.0011 | 0.0008 | 0.0014 | 0.0011 | 0.0008 | 0.0018 | 1.68 | 1.68 | 1.001 |

Table 8: The numerical results of the double differential cross section calculated with three different theoretical uncertainties for both the off-shell and on-shell state. The experimental data is based on [10]. Other information about the calculations is given at the beginning of section 5.

| | **All LO collisions** ||||||||||
| | **CDF ($\sqrt{s} = 1.96$ TeV , $|\eta^\gamma| < 1$)** ||||||||||
| $E_T^\gamma$ (GeV) | $\frac{d^2\sigma}{dE_T^\gamma d\eta^\gamma}\left(\frac{\text{pb}}{\text{GeV}}\right)$ (on-shell) ||| $\frac{d^2\sigma}{dE_T^\gamma d\eta^\gamma}\left(\frac{\text{pb}}{\text{GeV}}\right)$ (off-shell) ||| Exp. data | Data on off-shell Theory for $\mu_F = \mu_R = E_T^\lambda$ | Data on on-shell Theory for $\mu_F = \mu_R = E_T^\lambda$ | Off-shell on on-shell for $\mu_F = \mu_R = E_T^\lambda$ |
|---|---|---|---|---|---|---|---|---|---|---|
| | $\mu_F = \mu_R = 0.5 E_T^\lambda$ | $\mu_F = \mu_R = E_T^\lambda$ | $\mu_F = \mu_R = 2 E_T^\lambda$ | $\mu_F = \mu_R = 0.5 E_T^\lambda$ | $\mu_F = \mu_R = E_T^\lambda$ | $\mu_F = \mu_R = 2 E_T^\lambda$ | | | | |
| 31.7 | 24.25 | 21.76 | 19.46 | 25.62 | 22.99 | 20.56 | 123 | 5.35 | 5.65 | 1.056 |
| 36 | 14.29 | 12.64 | 11.19 | 14.99 | 13.27 | 11.74 | 62.1 | 4.68 | 4.91 | 1.049 |
| 41.1 | 8.19 | 7.15 | 6.27 | 8.54 | 7.45 | 6.53 | 31 | 4.16 | 4.33 | 1.042 |
| 46.5 | 4.85 | 4.19 | 3.65 | 5.01 | 4.33 | 3.77 | 17.2 | 3.97 | 4.1 | 1.033 |
| 53.8 | 2.61 | 2.22 | 1.92 | 2.67 | 2.28 | 1.97 | 7.93 | 3.47 | 3.56 | 1.026 |
| 63.9 | 1.24 | 1.05 | 0.90 | 1.26 | 1.06 | 0.91 | 3.54 | 3.33 | 3.37 | 1.013 |
| 74.1 | 0.654 | 0.55 | 0.46 | 0.66 | 0.55 | 0.47 | 1.76 | 3.19 | 3.21 | 1.007 |
| 84.1 | 0.377 | 0.31 | 0.26 | 0.38 | 0.31 | 0.26 | 0.91 | 2.89 | 2.9 | 1.003 |
| 97.2 | 0.1988 | 0.07 | 0.14 | 0.20 | 0.16 | 0.14 | 0.44 | 2.68 | 2.69 | 1.003 |
| 118 | 0.083 | 0.07 | 0.06 | 0.08 | 0.07 | 0.06 | 0.168 | 2.47 | 2.48 | 1 |
| 138 | 0.040 | 0.032 | 0.027 | 0.040 | 0.0324 | 0.027 | 7.25×10⁻² | 2.23 | 2.23 | 1.0005 |
| 158 | 0.0208 | 0.017 | 0.014 | 0.021 | 0.017 | 0.014 | 3.41×10⁻² | 2.04 | 2.03 | 1.0003 |
| 181 | 10.4×10⁻³ | 8.3×10⁻³ | 6.8×10⁻³ | 0.010 | 0.083 | 6.8×10⁻³ | 1.46×10⁻² | 1.76 | 1.75 | 1.001 |
| 212 | 4.4×10⁻³ | 3.4×10⁻³ | 2.8×10⁻³ | 4.4×10⁻³ | 3.5×10⁻² | 2.8×10⁻³ | 5.66×10⁻³ | 1.63 | 1.62 | 1 |
| 256 | 1.4×10⁻³ | 1.08×10⁻³ | 8.7×10⁻⁴ | 1.4×10⁻³ | 1.08×10⁻³ | 9×10⁻⁴ | 1.38×10⁻³ | 1.27 | 1.27 | 1.0002 |
| 350 | 1.3×10⁻³ | 1.01×10⁻³ | 8×10⁻⁵ | 1.3×10⁻⁴ | 1.01×10⁻³ | 8×10⁻⁵ | 1.49×10⁻⁴ | 1.46 | 1.46 | 1 |